\documentclass[a4paper]{article}

\usepackage{INTERSPEECH2021}
\usepackage{microtype}
\usepackage{graphicx}
\usepackage{amsmath}
\usepackage{amsfonts}
\usepackage{multirow}
\usepackage{booktabs}

\title{Speech Pre-training with Acoustic Piece}
\name{Shuo Ren$^1$, Shujie Liu$^1$, Yu Wu$^1$, Long Zhou$^1$, Furu Wei$^1$}
\address{
  $^1$Microsoft Research Asia}
\email{\{renshuo, shujliu, wu.yu, long.zhou, fuwei\}@microsoft.com}

\begin{document}
\maketitle
\begin{abstract}
Previous speech pre-training methods, such as wav2vec2.0 and HuBERT, pre-train a Transformer encoder to learn deep representations from audio data, with objectives predicting either elements from latent vector quantized space or pre-generated labels (known as target codes) with offline clustering. However, those training signals (quantized elements or codes) are independent across different tokens without considering their relations. According to our observation and analysis, the target codes share obvious patterns aligned with phonemized text data. Based on that, we propose to leverage those patterns to better pre-train the model considering the relations among the codes. The patterns we extracted, called \textbf{``acoustic piece''}s, are from the sentence piece result of HuBERT codes. With the acoustic piece as the training signal, we can implicitly bridge the input audio and natural language, which benefits audio-to-text tasks, such as automatic speech recognition (ASR). Simple but effective, our method \textbf{``HuBERT-AP''} significantly outperforms strong baselines on the LibriSpeech ASR task.
\end{abstract}

\section{Introduction}
Recent years have witness the emergence and success of speech pre-training methods, such as Wav2vec2.0 \cite{baevski2020wav2vec}, APC \cite{chung2020generative}, MPC \cite{jiang2021further}, and HuBERT \cite{hsu2021hubert}. Based on self-supervised training, those methods take large-scale audio data as the input and learn deep representations of them, benefiting the downstream speech tasks such as automatic speech recognition (ASR) and speaker identification (SID). Based on the Transfomer encoder architecture, HuBERT uses an offline clustering step to generate pseudo target labels with the MFCC features of input audio or the hidden representations of specific layers, and achieves better results compared with previous methods. The target labels are also called acoustic units or target codes. However, the HuBERT codes are generated one by one through the KNN, so the codes are independent among each other and the patterns inherent in it are not leveraged during pre-training. But according to our observation, the patterns show high relevance to phonemized text data and could be helpful in pre-training. 

Motivated by that, in this paper, we propose to leverage the patterns extracted from the target codes as the training signal for speech pre-training, to guide the model to learn better acoustic features. The proposed training target is called \textbf{``acoustic piece''}, which is built from the sentence piece results of the original HuBERT target codes. Because of its high relevance to phonemized natural language, the acoustic piece could significantly benefit audio-to-text tasks such ASR. We name our method \textbf{``HuBERT-AP''}, meaning the \textbf{HuBERT} model pre-trained with the \textbf{A}coustic \textbf{P}iece. Experimental results on the LibriSpeech ASR task show the effectiveness and superiority of our method. Our contributions are listed as follows:

\begin{itemize}
    \item We analyze HuBERT codes and find patterns in them, which show high relevance to natural language.
    \item We extract the patterns in HuBERT codes, named ``acoustic piece'', and take it as the target label for speech pre-training.
    \item Our method is very simple but effective, significantly outperforming previous strong baselines on the LibriSpeech ASR task.
\end{itemize}

\section{Prerequisite}
\subsection{HuBERT}
\begin{figure}[t]
	\centering
	\includegraphics[width=7.5cm]{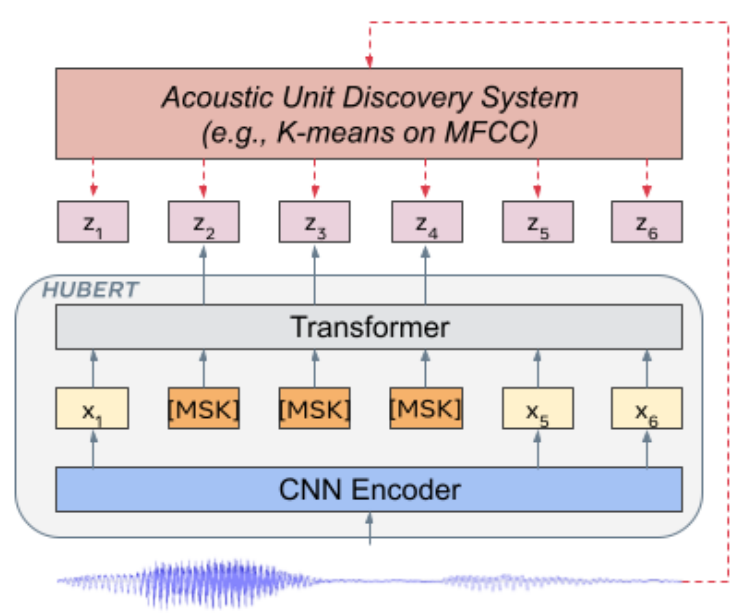}
	\caption{The HuBERT model \cite{hsu2021hubert}.
	}\label{fig:hubert}
\end{figure}

The Hidden-Unit BERT (HuBERT) \cite{hsu2021hubert} is an approach for self-supervised speech representation learning, which utilizes an offline clustering step to provide aligned target labels (codes) for a BERT-like prediction loss. Speech signals differ from the text in that they are continuous-valued sequences. Therefore, in their method, they use the offline clustering to generate discrete labels for a BERT-like pre-training. The generated target labels are also called \textbf{``acoustic units''} or target codes. The whole architecture of HuBERT is shown in Figure \ref{fig:hubert}. During pre-training, the model predicts hidden cluster assignments of the masked frames ($z_2$, $z_3$, $z_4$ in the figure) generated by one or more iterations of k-means clustering. The HuBERT base model performs two iterations. In the first iteration, the assigned labels are generated with k-means clustering (k=100) on the MFCC features extracted from the raw audio data. In the second iteration, the labels are generated with k-means clustering (k=500) based on the 6th layer hidden representations of the HuBERT model after the first iteration. 

\subsection{Analysis on HuBERT Codes}
The HuBERT codes are shown to be in good quality of phone purity, cluster purity and phone-normalized mutual information according to \cite{hsu2021hubert}. Frames of the same phonemes are more likely labeled as similar sequences of codes. In this subsection, we give some examples chosen from LibriSpeech 960h dataset to show the high relevance of HuBERT codes to text data and the patterns inherent in them. 

\begin{figure}[t]
	\centering
	\includegraphics[width=7.5cm]{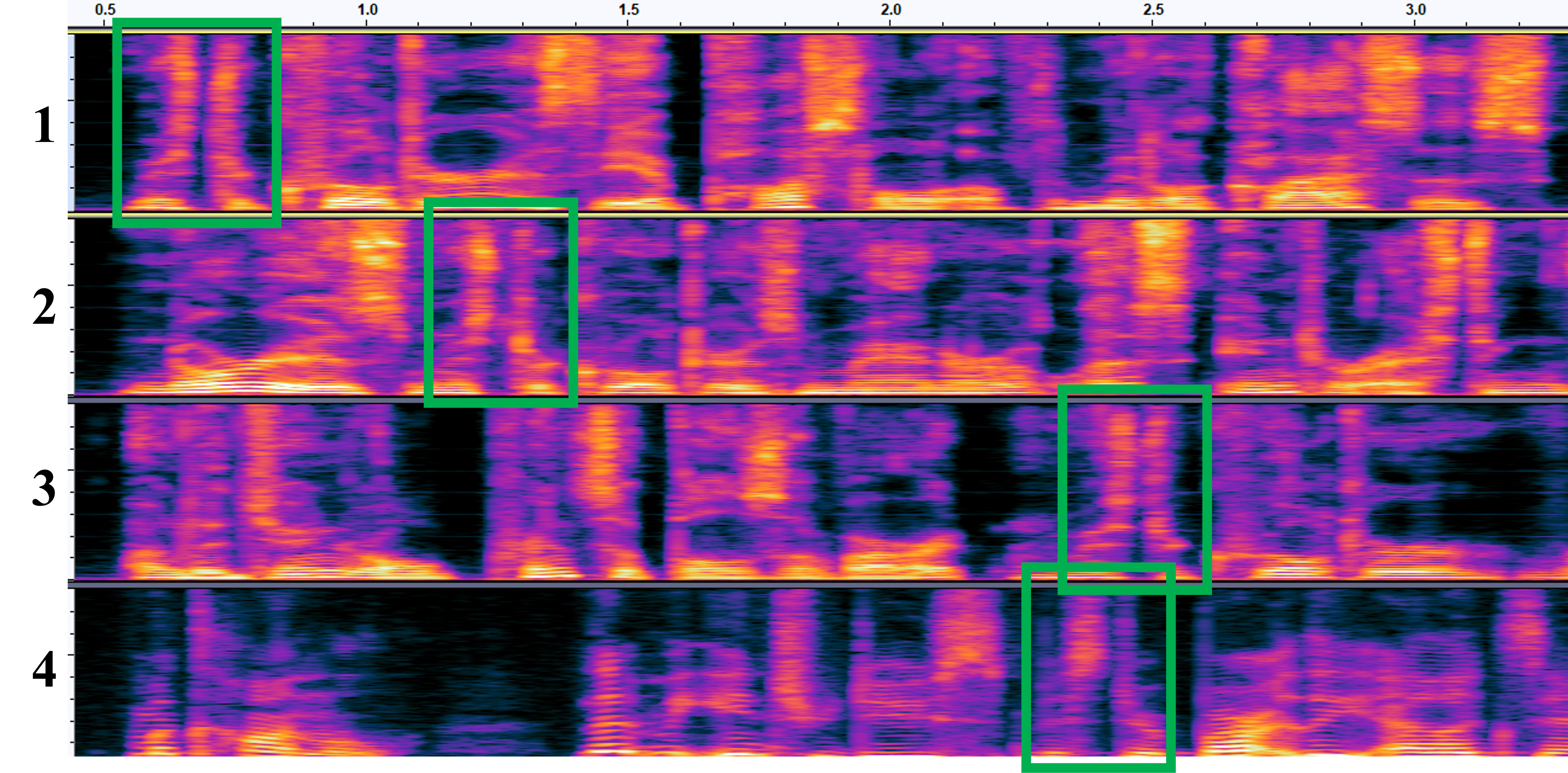}
	\caption{The Mel-Spectrum of four sentences. The part in the green box corresponds to the word ``mister''.
	}\label{fig:mister}
\end{figure}

\begin{table}[t]
\begin{center}
\small
\begin{tabular}{l|c}
    \toprule
     & Codes corresponding to ``mister'' in the four samples  \\
    \midrule
    1&...178 \textbf{285} \textbf{285} \textbf{285} \textbf{285} \textbf{279} \textbf{279} \textbf{138} \textbf{374} \textbf{374} \textbf{374} 224...  \\
    \hline
    2&...309 \textbf{285} \textbf{285} \textbf{285} \textbf{285} 378 \textbf{279} \textbf{138} \textbf{374} \textbf{374} \textbf{374} 52... \\
    \hline
     3&...178 \textbf{285} \textbf{285} \textbf{285} \textbf{285} 378 \textbf{279} \textbf{138} \textbf{374} \textbf{374} 52... \\
    \hline
     4&...309 \textbf{285} \textbf{285} \textbf{285} \textbf{285} 258 378 \textbf{279} \textbf{138} \textbf{374} \textbf{374} 52... \\
    \bottomrule
\end{tabular}
\end{center}
\caption{\label{tab:mister} The codes of ``mister'' in different samples. The bold codes are shared by the four samples.}
\end{table}

Figure \ref{fig:mister} shows the Mel-Spectrum of four samples in the dataset containing the word ``mister'', and we use the green box to show the part corresponding to the word (human-labeled). The first three samples are man utterances while the last one is a woman utterance. With no surprise, the Mel-Spectrum of the same word in different sentences looks similar. Then we observe the HuBERT codes corresponding to the parts and list them in Table \ref{tab:mister}. The codes are generated with k-means clustering on the 6th-layer representations of the released pre-trained HuBERT base model \footnote{https://github.com/pytorch/fairseq/tree/main/examples/hubert}. In this table, we find that the codes corresponding to ``mister'' in different samples are also similar and show an obvious pattern. The codes they share are ``285 279 138 374''; some codes such as ``178'', ``309'', ``378'', ``52'' are shared by some of them. There are also some unique codes in the samples, such as ``224'' in the first sample and ``258'' in the last samples, which may be noisy labels. Apart from the word ``mister'', we also analyze other examples (such as ``the'', ``and'' and ``child'') and find the similar observation. With clear and obvious patterns inherent in the codes, we propose to leverage these patterns to better guide HuBERT pre-training with context-dependent information, benefiting the audio-to-text tasks such as ASR.

\begin{figure}[t]
	\centering
	\includegraphics[width=7.5cm]{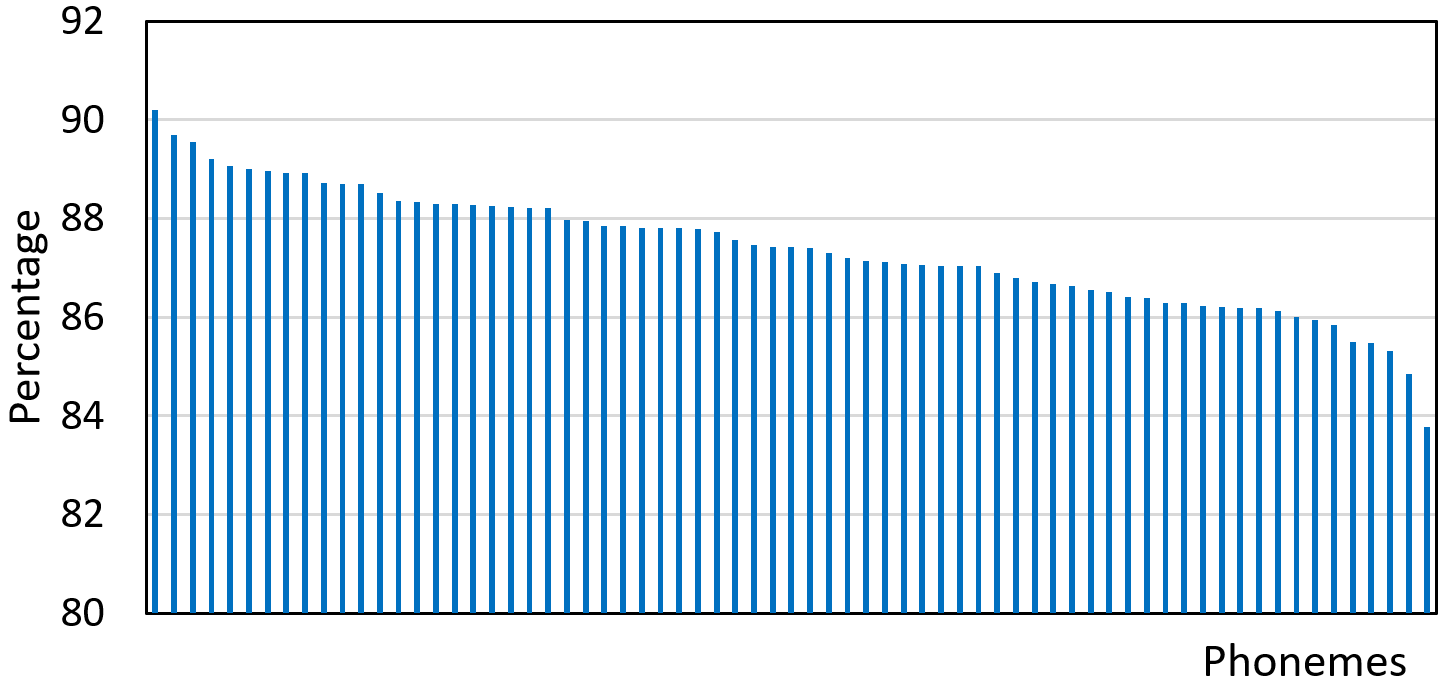}
	\caption{The average percentage of the sharing code for each phoneme. The x-axis denote 69 phonemes sorted according to the value of the y-axis.
	}\label{fig:percentage}
\end{figure}

Apart from the example above, we also conduct a quantitative analysis on the HuBERT codes. We first do force alignment on the LibriSpeech 100h data and get the alignment information between the audio and the transcript on phoneme level. Next, we align the HuBERT codes with the phonemized text according to the aligned interval and the sampling rate. Then, we collect the shared codes aligned with each phoneme and record them as the total set. Finally, we calculate the average percentage of the sharing code in the total set of each phoneme. The result is shown in Figure \ref{fig:percentage}. From the figure, we see that most phonemes sharing over 86\% codes, indicating that there are obvious patterns in the codes aligned with phonemes.

\section{Method}
\begin{figure}[t]
	\centering
	\includegraphics[width=7.5cm]{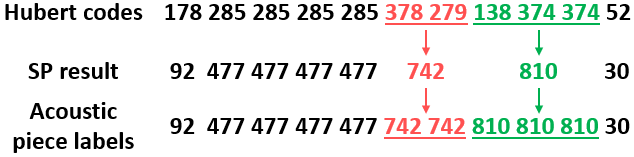}
	\caption{Example of the acoustic piece generation.
	}\label{fig:acoustic_piece}
\end{figure}

Instead of extracting complex hierarchical patterns as training signals, in this paper, we leverage sentence piece \cite{kudo2018sentencepiece} on the codes to automatically merge the highly frequent code patterns into one piece. We call the merged piece ``acoustic piece'', following ``acoustic units'' in the HuBERT paper. We use the acoustic piece label as the training target for pre-training. Specifically, we do the sentence piece on the original HuBERT codes, with pre-defined vocabulary sizes of 1k, 2k and 3k, bigger than the number of the original HuBERT target label, aka 500.  The sentence piece operation will merge some codes into one acoustic piece, and we then assign each merged code with the same piece id, to keep the length of the target label unchanged. As shown in Figure \ref{fig:acoustic_piece}, the first line is the original HuBERT codes, and the second line is the sentence piece (SP) result with the vocabulary size 1k. We find that the code segment ``378 279'' is merged into one piece `` 742'' and ``138 374 374'' is merged into the new piece ``810''. There is no merge operation on the remaining codes but they are all assigned with the new id in the acoustic piece vocabulary. After that, we re-assigned each merged code with their new acoustic piece id and keep the length of the target label unchanged, as the third line shows. 

Our model is based on the HuBERT base model, which is a 12-layer Transformer encoder with a 7-layer CNN pre-net extracting the features from raw audio data. Following WavLM \cite{chen2021wavlm}, we also employ gated relative position bias \cite{chi2021xlm} in our model, which is encoded based on the offset between the “key” and “query” in the Transformer self-attention mechanism. Taking the unlabeled raw audio data as the input, the final encoder layer is used to predict the target labels and calculate the cross-entropy loss. Leveraging the ASR audio-text paired data, we fine-tune the pre-trained model with the supervised CTC loss \cite{graves2006connectionist}. During inference, we use the wav2letter++ \cite{pratap2019wav2letter++} beam search decoder for language model-fused decoding, which optimizes:
\begin{equation}
    \log P_{CTC}(Y|X)+w_1 \log P_{LM}(Y) +w_2|Y|
\end{equation}
where $Y$ is the predicted text, $|Y|$ is the length of the text, and $w_1$ and $w_2$ denote the language model weight and word score. The decoding hyperparameters are searched with Ax\footnote{https://github.com/facebook/Ax}.

\section{Experiment}
\subsection{Setup}
\subsubsection{Dataset}
For pre-training, we leverage large-scale unlabeled English audio data from diverse domains. Following \cite{chen2021wavlm}, we collect 94k hours of data in total, including Libri-Light \cite{kahn2020libri}, VoxPopuli \cite{wang2021voxpopuli}, and GigaSpeech \cite{chen2021gigaspeech}. All the data is used for pre-training our base model. Libri-Light is derived from open-source audiobooks from the LibriVox project. It contains over 60k hours of audio. VoxPopuli is a large-scale multi-lingual unlabeled audio dataset, collected from 2009-2020 European Parliament event recordings. We use 24k hours of English data of it for pre-training. Gigaspeech is collected from audiobooks, podcasts and YouTube with a variety of topics \footnote{GigaSpeech contains 40k hours speech data in total, with 30k not well processed. Therefore we only use 10k of it.}.

For the ASR fine-tuning, we consider three different partitions: the train clean-100 subset (100 hours labeled data) of LibriSpeech \cite{panayotov2015librispeech}, as well as the Libri-Light \cite{kahn2020libri} limited resource training subsets originally extracted from LibriSpeech, including train-1h (1 hours labeled data) and train-10h (10 hour labeled data). We follow the evaluation protocol of Libri-Light for these splits and evaluate on the standard LibriSpeech test-clean/other sets.

\subsubsection{Implementation Details}
We implement and pre-train our models following previous work \cite{hsu2021hubert}\footnote{https://github.com/pytorch/fairseq/tree/main/examples/hubert}. The model architecture is the same as HuBERT base model, which contains 12 encoder layers, 768-dimensional hidden states, and 8 attention heads, resulting in 94.70M parameters. The relative position embedding is shared by all layers, so it does not significantly increase the number of parameters. We first use the 6th layer of the released HuBERT model to generate original labels with the offline clustering, then do sentence piece on it with different vocabulary to generate acoustic piece labels. We pre-train the model for 400k steps with the Adam optimizer \cite{kingma2014adam}. The learning rate ramps up linearly from 0 to the peak learning rate of 5e-4 for the first 8\% of the training steps, and then decays linearly back to 0. The CNN audio pre-net is also the same as in HuBERT, which generates a feature sequence at a 20ms framerate for audio sampled at 16kHz (CNN down-sampling factor is 320x). We use the public code for sentence piece\footnote{https://github.com/google/sentencepiece}. 

\subsubsection{Fine-tuning Setup}
During fine-tuning we apply a masking strategy to the feature encoder outputs similar to SpecAugment \cite{park2019specaugment}: we randomly choose a number of starting time steps for which a span of ten subsequent time-steps is replaced with a mask embedding; spans may overlap and we use the same masked time step embedding as during pre-training. We also mask channels by choosing a number of channels as starting indices and then expand each one to cover the subsequent 64 channels. Spans may overlap and the selected channel spans are set to zero value. We use LayerDrop \cite{huang2016deep,fan2019reducing} at a rate of 0.05 during fine-tuning. The fine-tuning hyper-parameter settings such as the time-step mask probability, channel mask probability and the total update steps used for different labeled data are the same as \cite{chen2021wavlm}. Table \ref{tab:decode_hyper} shows the decoding parameters used for final evaluations of the various labeled data setups.



\begin{table}[!t]
\begin{center}
\small
\begin{tabular}{@{\extracolsep{10pt}}lcc}
    \toprule
     & 4-gram LM weight & 4-gram word insert \\
    \midrule
     1 hour & 3.09 & -2.33 \\
     10 hour & 2.12 & -0.90 \\
     100 hour & 2.15 & -0.52 \\
    \bottomrule
\end{tabular}
\end{center}
\caption{\label{tab:decode_hyper} Decoding parameters for Librispeech subsets.}
\end{table}

\begin{table}[!h]
\begin{center}
\small
\begin{tabular}{@{\extracolsep{10pt}}lccc@{}}
    \toprule
     \multirow{2}{*}{Model} & \multirow{2}{*}{LM} & \multicolumn{2}{c}{test}\\ \cline{3-4}
    &&clean&other \\
    \midrule\midrule
    \multicolumn{4}{c}{\textbf{\textit{1-hour labeled}}} \\
    \midrule
    wav2vec 2.0 \textsc{Base} & None & 24.5 & 29.7 \\
    WavLM \textsc{Base} & None & 24.5 & 29.2 \\
    WavLM \textsc{Base+} & None & 22.8 & 26.7 \\
    *HuBERT-AP \textsc{Base} & None & 17.0 & 23.3 \\
    *HuBERT-AP \textsc{Base+} & None & 16.9 & 22.3 \\
    \hline
    DeCoAR 2.0  & 4-gram &  13.8 & 29.1 \\
    DiscreteBERT  & 4-gram & 9.0 & 17.6 \\
    wav2vec 2.0 \textsc{Base} & 4-gram & 5.5 & 11.3 \\
    HuBERT \textsc{Base} & 4-gram & 6.1 & 11.3 \\
    WavLM \textsc{Base} & 4-gram & 5.7 & 10.8 \\
    WavLM \textsc{Base+} & 4-gram & 5.4 & 9.8 \\
    *HuBERT-AP \textsc{Base} & 4-gram & 5.5 & 10.6 \\
    *HuBERT-AP \textsc{Base+} & 4-gram & 5.3 & 9.6 \\
    \midrule
    \multicolumn{4}{c}{\textbf{\textit{10-hour labeled}}} \\
    \midrule
    wav2vec 2.0 \textsc{Base} & None & 11.1 & 17.6 \\
    WavLM \textsc{Base} & None & 9.8 & 16.0 \\
    WavLM \textsc{Base+} & None & 9.0 & 14.7 \\
    *HuBERT-AP \textsc{Base} & None & 9.1 & 15.2 \\
    *HuBERT-AP \textsc{Base+} & None & 8.4 & 13.9 \\
    \hline
    DeCoAR 2.0  & 4-gram &  5.4 & 13.3 \\
    DiscreteBERT  & 4-gram & 5.9 & 14.1 \\
    wav2vec 2.0 \textsc{Base} & 4-gram & 4.3 & 9.5 \\
    HuBERT \textsc{Base} & 4-gram & 4.3 & 9.4 \\
    WavLM \textsc{Base} & 4-gram & 4.3 & 9.2 \\
    WavLM \textsc{Base+} & 4-gram & 4.2 & 8.8 \\
    *HuBERT-AP \textsc{Base} & 4-gram & 4.2 & 9.0  \\
    *HuBERT-AP \textsc{Base+} & 4-gram & 4.1 & 8.4 \\
    \midrule
    \multicolumn{4}{c}{\textbf{\textit{100-hour labeled}}} \\
    \midrule
    wav2vec 2.0 \textsc{Base} & None & 6.1 & 13.3 \\
    WavLM \textsc{Base} & None & 5.7 & 12.0 \\
    WavLM \textsc{Base+} & None & 4.6 & 10.1 \\
    *HuBERT-AP \textsc{Base} & None & 4.9 & 10.7 \\
    *HuBERT-AP \textsc{Base+} & None & 4.6 & 9.5 \\
    \hline
    DeCoAR 2.0  & 4-gram &  5.0 & 12.1 \\
    DiscreteBERT  & 4-gram & 4.5 & 12.1 \\
    wav2vec 2.0 \textsc{Base} & 4-gram & 3.4 & 8.0 \\
    HuBERT \textsc{Base} & 4-gram & 3.4 & 8.1 \\
    WavLM \textsc{Base} & 4-gram & 3.4 & 7.7 \\
    WavLM \textsc{Base+} & 4-gram & 2.9 & 6.8 \\
    *HuBERT-AP \textsc{Base} & 4-gram & 3.1 & 7.1 \\
    *HuBERT-AP \textsc{Base+} & 4-gram & 2.9 & 6.6 \\
    \bottomrule
\end{tabular}
\end{center}
\caption{\label{tab:main_result} WER of ASR on the LibriSpeech and test sets, when trained on the LibriLight low-resource labeled data setups of 1 hour, 10 hours and the clean 100h subset of LibriSpeech. * means our method. + means pre-training with 1M update steps.}
\end{table}

\subsection{Main Results}
We compare our method with several competitive self-supervised approaches in the literature, including DeCoAR 2.0 \cite{ling2020decoar}, DiscreteBERT \cite{baevski2020effectiveness}, wav2vec 2.0 \cite{baevski2020wav2vec}, HuBERT \cite{hsu2021hubert} and WavLM \cite{chen2021wavlm}. The results are shown in Table \ref{tab:main_result}. Due to time limitations, we only compare the base model of each method and use the 4-gram language model for LM fusion. From the table, without LM fusion, our method HuBERT-AP outperforms the previous methods by a large margin for all fine-tuning splits with an improvement of about 5\% to 20\%. With LM fusion, our method still outperforms most previous baselines and on par with the strong baseline WavLM.

It seems unfair to directly compare our method with the HuBERT base model because our model is pre-trained with larger dataset than HuBERT. Therefore, we also pre-train our model with only LibriSpeech 960h data, which improves only 3\% percent compared with the HuBERT baseline after fine-tuning. This may be because the extracted patterns are not learned well with limited pre-training data. Actually, the fair baselines are the WavLM models, which uses the same pre-training data as ours. From the table, we see that our models achieve significantly better results than WavLM, showing the effectiveness of our method.  

\subsection{Analysis of Acoustic Piece}
To investigate the superiority of the acoustic piece, we conduct a quantitative analysis to evaluate the segmentation quality of acoustic pieces. We use the force alignment result on the dev clean set of LibriSpeech to find the golden phoneme boundaries. We compare the boundary in the original HuBERT code sequence as well as the corresponding acoustic pieces generated with our method, to check whether the code/piece boundaries are consistent with the phoneme boundaries. We report precision, recall and F-measure in Table \ref{tab:boundary}. From the table, we see that the acoustic piece is of higher quality and predicts more accurate phoneme boundaries, which can benefit the pre-training. 

\begin{table}[!htp]
\begin{center}
\small
\begin{tabular}{l|ccc}
    \toprule
    Method & Precision & Recall & F1  \\
    \midrule
    HuBERT codes & 0.387 & 0.672 & 0.421  \\
    Acoustic piece & 0.579 & 0.712 & 0.628 \\
    \bottomrule
\end{tabular}
\end{center}
\caption{\label{tab:boundary} Quantitative evaluation of segment boundaries of different methods wrt. golden phoneme boundaries.}
\end{table}

\begin{table}[!htp]
\begin{center}
\small
\begin{tabular}{@{\extracolsep{10pt}}lcc@{}}
    \toprule
    Model & test clean & test other \\
    \midrule
     HuBERT & 3.4 & 8.1 \\
     AP 1k  & \textbf{3.1} & \textbf{7.1} \\
     AP 1k + HuBERT  & 3.2 & 7.1 \\
     AP 2k  & 3.2 & 7.3  \\
     AP 3k  & 3.1 & 7.2 \\
     AP 5k & 3.2 & 7.3 \\
     AP 10k & 3.3 & 7.5 \\
    \bottomrule
\end{tabular}
\end{center}
\caption{\label{tab:ablation_vocab} The influence of vocabulary size of the sentence piece model. ``AP'' means our method and ``1k'', ``2k'' ,``3k'', ``5k'', ``10k'' mean different vocabulary sizes.}
\end{table}

\subsection{Ablation Study}
To investigate the influence of the vocabulary size of the acoustic pieces, we conduct the ablation study with the 100-hour fine-tuning setting with the 4-gram language model fusion and show the results in Table \ref{tab:ablation_vocab}. From the table, we see that our method is robust to the sentence piece vocabulary size (the number of the piece units). With the increase of the piece units, the ASR fine-tuning results remain nearly unchanged, even get a little worse than 1k. This may result from the sparsity brought by the large vocabulary size. We also investigate the influence of pre-training with the combined labels, i.e. the HuBERT labels and the proposed acoustic piece labels (``AP 1k + HuBERT''). The combined labels do not improve the performance compared with the sole acoustic piece labels. Therefore, we recommend using the 1k acoustic piece units setting by default because of its better performance and lower memory cost during pre-training. 

In addition, we also conduct experiments to test the influence of the gated relative position bias adopted in our method. We choose the settings of fine-tuning with 100h labeled data. The result is shown in table \ref{tab:ablation_relpos}. From the table, we find that relative position bias improves the performance without LM fusion but takes no effect with LM fusion compared with the HuBERT baseline. In contrast, our method HuBERT-AP consistently improve the performance over the two baselines. 

\begin{table}[!htp]
\begin{center}
\small
\begin{tabular}{@{\extracolsep{10pt}}lccc@{}}
    \toprule
     \multirow{2}{*}{Model} & \multirow{2}{*}{LM} & \multicolumn{2}{c}{test}\\ \cline{3-4}
    &&clean&other \\
    \midrule
    HuBERT & None & 6.0 & 13.1 \\
    HuBERT+relpos & None & 5.7 & 12.3 \\
    *HuBERT-AP & None & 4.9 & 10.7 \\
    \hline
    HuBERT & 4-gram & 3.4 & 8.1 \\
    HuBERT+relpos & 4-gram & 3.4 & 8.1 \\
    *HuBERT-AP & 4-gram & 3.1 & 7.1 \\
    \bottomrule
\end{tabular}
\end{center}
\caption{\label{tab:ablation_relpos} The influence of the relative position bias (HuBERT+relpos). * means our method. }
\end{table}

\section{Conclusion}
In this paper, we propose a simple but effective speech pre-training method named HuBERT-AP. We first analyze the HuBERT codes and find there are patterns that are highly relevant to natural language. Based on that, we extract the patterns called ``acoustic piece'' with the sentence piece method, and take it as the training signal for speech pre-training. Experiments show that HuBERT-AP consistently improves the ASR performance on different settings and outperforms strong baselines. In the future, we will extend our method to large model with more pre-training data to further verify the effectiveness of our method.

\bibliographystyle{IEEEtran}

\bibliography{mybib}

\clearpage

\end{document}